\documentclass[11pt,draftcls,onecolumn]{IEEEtran}

\usepackage{graphicx}

\usepackage[usenames]{color}
 \usepackage{verbatim}

\usepackage{latexsym}
\usepackage{amstext,amssymb, amsfonts, amsthm}
\usepackage[cmex10]{amsmath}

\long\def\symbolfootnote[#1]#2{\begingroup%
\def\thefootnote{\fnsymbol{footnote}}\footnote[#1]{#2}\endgroup}

\def\Prob{{\sf P}}

\def\Expect{{\sf E}}

\def\sq{\hbox{\rlap{$\sqcap$}$\sqcup$}}
\def\qed{\ifmmode\sq\else{\unskip\nobreak\hfil
\penalty50\hskip1em\null\nobreak\hfil\sq
\parfillskip=0pt\finalhyphendemerits=0\endgraf}\fi\medskip}

 \def\FRAC#1#2#3{\genfrac{}{}{}{#1}{#2}{#3}}

\def\ddtp{{\mathchoice{\FRAC{1}{d^{\hbox to 2pt{\rm\tiny +\hss}}}{dt}}%
{\FRAC{1}{d^{\hbox to 2pt{\rm\tiny +\hss}}}{dt}}%
{\FRAC{3}{d^{\hbox to 2pt{\rm\tiny +\hss}}}{dt}}%
{\FRAC{3}{d^{\hbox to 2pt{\rm\tiny +\hss}}}{dt}}}}

\newtheorem{theorem}{Theorem}[section]

\newtheorem{lemma}[theorem]{Lemma}

\def\eq#1/{(\ref{e:#1})}

\newcommand{\field}[1]{\mathbb{#1}}

\def\Re{\field{R}}

\newcounter{rmnum}
\newenvironment{romannum}{\begin{list}{{\upshape (\roman{rmnum})}}{\usecounter{rmnum}
\setlength{\leftmargin}{10pt} \setlength{\rightmargin}{8pt}
\setlength{\itemindent}{-1pt} }}{\end{list}}

\newcounter{anum}

\def\til={{\widetilde =}}

\def\clF{{\cal F}}

\def\clN{{\cal N}}

\def\clP{{\cal P}}

\def\clX{{\cal X}}

\newlength{\noteWidth}
 \long\def\notes#1{\ifinner
             {\tiny #1}
             \else
             \marginpar{\parbox[t]{\noteWidth}{\raggedright\tiny #1}}
             \fi}

\def\notes#1{}

\def\ofp{(\Omega, \clF, \mathsf{P})}
\def\Prob{\mathsf{P}}
\def\esssup{\text{ ess sup }}
\def\jsrp{J_{\scriptscriptstyle \mathrm{SRP}}}
\def\tsrp{\tau_{\scriptscriptstyle \mathrm{SRP}}}
\long\def\notes#1{\ifinner
             {\tiny #1}
             \else
              \marginpar{\parbox[t]{\noteWidth}{\raggedright\tiny #1}}
               \fi}

\begin{document}

\title{Minimax Robust Quickest Change Detection}

\author{
\authorblockN{Jayakrishnan Unnikrishnan, Venugopal~V.~Veeravalli, Sean
Meyn}\\
\authorblockA{Department of Electrical and Computer Engineering, and Coordinated Science Laboratory\\
University of Illinois at Urbana-Champaign}\footnote[1]{The
authors are with the Department of Electrical and Computer
Engineering and the Coordinated Science Laboratory, University
of Illinois at Urbana-Champaign, Urbana, IL. Email: \{junnikr2,
vvv, meyn\}@illinois.edu.

This work was supported by NSF Grants CCF 07-29031 and CCF
08-30169, through the University of Illinois. Portions of the
results presented here were published in abridged form in
\cite{unnveemey09}. } }

\maketitle


\begin{abstract}
The popular criteria of optimality for quickest change
detection procedures are the Lorden criterion, the
Shiryaev-Roberts-Pollak criterion, and the Bayesian criterion.
In this paper a robust version of these quickest change
detection problems is considered when the pre-change and
post-change distributions are not known exactly but belong to
known uncertainty classes of distributions. For uncertainty
classes that satisfy a specific condition, it is shown that one
can identify least favorable distributions (LFDs) from the
uncertainty classes, such that the detection rule designed for
the LFDs is optimal for the robust problem in a minimax sense.
The condition is similar to that required for the
identification of LFDs for the robust hypothesis testing
problem originally studied by Huber. An upper bound on the
delay incurred by the robust test is also obtained in the
asymptotic setting under the Lorden criterion of optimality.
This bound quantifies the delay penalty incurred to guarantee
robustness. When the LFDs can be identified, the proposed test
is easier to implement than the CUSUM test based on the
Generalized Likelihood Ratio (GLR) statistic which is a popular
approach for such robust change detection problems. The
proposed test is also shown to give better performance than the
GLR test in simulations for some parameter values.
\end{abstract}


\section{Introduction}
The problem of detecting an abrupt change in a system based on
observations is a dynamic hypothesis testing problem with a
rich set of applications. Such problems of change detection
were first studied by Page over fifty years ago in the context
of quality control \cite{page54}. In its standard formulation
there is a sequence of observations whose distribution changes
at some unknown point in time, referred to as the
`change-point'. The goal is to detect this change as soon as
possible, subject to a false alarm constraint. Some
applications of change detection are intrusion detection in
computer networks and security systems, detecting faults in
infrastructure of various kinds, and spectrum monitoring for
opportunistic access to wireless networks.


Most of the past work in the area of change detection has been
restricted to the setting where the distributions of the
observations prior to the change and after the change are known
exactly (see, e.g., \cite{mous86}, \cite{lorden71},
\cite{pollak85}, \cite{shiry78}; for an overview of the work in
this area, see \cite{tart91}, \cite{basnik93} and
\cite{brodar93}.). The three most popular criteria for
optimizing the tradeoff between detection delay and false alarm
rate are the Lorden criterion \cite{lorden71} and the
Shiryaev-Roberts-Pollak criterion, in which the change-point is
a deterministic quantity, and Shiryaev's Bayesian formulation
\cite{shiry63}, in which the change-point is modeled as a
random variable with a known prior distribution. In this paper
we study all these three versions of change detection, under
the setting where the pre-change and post-change distributions
are not known exactly but belong to known uncertainty classes.
We pose a minimax robust version of the standard quickest
change detection problem wherein the objective is to identify
the change detection rule that minimizes the maximum delay over
all possible distributions. This minimization should be
performed while meeting the false alarm constraint for all
possible values of the unknown distributions. We obtain a
solution to this problem when the uncertainty classes satisfy
some specific conditions. Under these conditions we can
identify Least Favorable Distributions (LFDs) from the
uncertainty classes, and the optimal robust change detection
rule is then the optimal (non-robust) change detection rule for
the LFDs. These conditions are similar to those given by Huber
\cite{huber65} for robust hypothesis testing problems. We also
discuss related results on robust sequential detection
\cite{huber65} \cite{qua85} later in the paper.

Although there has been some prior work on robust change
detection, these approaches are distinctly different from ours.
The maximin approach of \cite{crosch94} is similar in that they
also identify LFDs for the robust problem. However, their
result is restricted to asymptotic optimality (as the false
alarm constraint goes to zero) under the Lorden criterion. A
similar formulation is also discussed in
\cite[Sec.7.3.1]{poorhadji09}. Some other approaches to this
problem (e.g. \cite{tartpol08}, \cite{gorpol95}) are aimed at
developing algorithms for quickest change detection with
unknown distributions. These works study the asymptotic
performance of the proposed tests under different distributions
but do not seek to guarantee minimax robustness over a given
class of distributions.

A closely related problem is the composite quickest change
detection problem. In general, these problems also address the
setting where the pre-change and post-change distributions are
unknown. However, unlike the robust problem, in composite
problems one seeks to identify a change detection procedure
that is simultaneously optimal under all possible values of the
unknown distributions. Exact solutions to these problems are
often intractable and hence most results are restricted to
asymptotic optimality. One such solution to a composite change
detection problem is discussed in \cite{lorden71} when only the
post-change distribution is unknown. In \cite{lorden71} a test
is given that is asymptotically optimal under the Lorden
criterion for all possible values of the unknown post-change
distribution in a one-dimensional exponential family of
distributions. This test is also referred to as the Generalized
Likelihood Ratio Test (GLR Test), and was also studied in
\cite{lai01} and \cite{sieven95}. An alternate asymptotically
optimal solution for the setting in which both pre-change and
post-change distributions are unknown was studied in
\cite{brodar05}.

We provide a performance comparison of our proposed robust test
with the GLR test. Although the GLR test asymptotically
performs as well as the optimal test with known distributions,
we show via simulations that our robust test can give improved
performance over the GLR test for moderate values of the false
alarm constraint. The GLR test is also often prohibitively
complex to implement in practice, while the proposed robust
CUSUM test admits a simple recursive implementation.

For the asymptotic version of the problem, we also provide an
analytical upper bound on the delay incurred by our robust test
and use it to provide an upper bound on the drop in performance
of our test relative to the optimal non-robust test.


The rest of the paper is organized as follows. We first state
the problem that we are studying in Section
\ref{sec:probstate}. In Section \ref{sec:robqcd} we describe
the robust solution and present some analysis. We discuss some
examples in Section \ref{sec:example} and conclude in Section
\ref{sec:conclusion}.

\section{Problem Statement} \label{sec:probstate}
In the online quickest change detection problem we are given
observations from a sequence $\{X_n: i = 1,2,\ldots\}$ taking
values in a set $\clX$. There are two known distributions
$\nu_0, \nu_1 \in \clP(\clX)$ where $\clP(\clX)$ is the set of
probability distributions on $\clX$. Initially, the
observations are drawn i.i.d. under distribution $\nu_0$. Their
distribution switches abruptly to $\nu_1$ at some unknown time
$\lambda$ so that $X_n \sim \nu_0$ for $n \leq \lambda-1$ and
$X_n \sim \nu_1$ for $n \geq \lambda$. The observations are
stochastically independent conditioned on the change-point. The
objective is to identify the occurrence of change with minimum
delay subject to false alarm constraints. We use
$\Expect_m^\nu$ to denote the expectation operator and
$\mathsf{P}_m^\nu$ to denote the probability law when the
change happens at $m$ and the pre-change and post-change
distributions are $\nu_0$ and $\nu_1$ respectively. The symbols
are replaced with $\Expect_\infty^\nu$ and
$\mathsf{P}_\infty^\nu$ when the change does not happen.
Similarly, if the pre-change and post-change distributions are
some $\mu$ and $\gamma$, respectively, and the change happens
at time $m$, we use $\Expect_m^{\mu, \gamma}$ to denote the
expectation operator and $\mathsf{P}_m^{\mu, \gamma}$ the
probability law. We further use $\clF_m$ to denote the sigma
algebra generated by $(X_1, X_2, \ldots, X_m)$.


A sequential change detection procedure is characterized by a
stopping time $\tau$ with respect to the observation sequence.
The design of the quickest change detection procedure involves
optimizing the tradeoff between two performance measures:
detection delay and frequency of false alarms. There are
various standard mathematical formulations for the optimal
tradeoff. In the minimax formulation of \cite{lorden71} the
change-point is assumed to be an unknown deterministic
quantity. The worst-case detection delay is defined as,
\[
\text{WDD}(\tau) = \sup_{\lambda \geq 1} \esssup \Expect_\lambda^\nu[ (\tau - \lambda + 1)^+ | \clF_{\lambda-1} ]
\]
where $x^+ = \max(x,0)$. This quantity captures the worst-case
value of the expected detection delay over all possible
locations of the change-point and all possible realizations of
the pre-change observations. The false alarm rate is defined
as,
\[
\text{FAR}(\tau) = \frac{1}{\Expect_\infty^\nu [\tau]}.
\]
Here $\Expect_\infty^\nu [\tau]$ can be interpreted as the mean
time to false alarm. Under the Lorden criterion, the objective
is to find the stopping rule that minimizes the worst-case
delay subject to an upper bound on the false alarm rate:
\begin{equation} \label{eqn: lordencriterion}
\text{Minimize} \text{ WDD}(\tau) \text{ subject to } \text{FAR}(\tau) \leq \alpha
\end{equation}
It was shown by Moustakides \cite{mous86}  that the optimal
solution to (\ref{eqn: lordencriterion}) is given by the
cumulative sum (CUSUM) test proposed by Page \cite{page54}. We
describe this test later in the paper.

An alternate formulation of the change detection problem was
studied by Pollak \cite{pollak85}. Even here the change point
is modeled as a deterministic quantity. But the delay to be
minimized is no longer the worst-case delay but a worst-case
average delay (also referred to as supremum average detection
delay by some authors) defined by,
\[
\jsrp(\tau) = \sup_{\lambda \geq 1} \Expect_\lambda^\nu[ \tau - \lambda | \tau \geq \lambda ].
\]
The Shiryaev-Roberts-Pollak criterion of optimality of a stopping rule $\tau$ for change detection is given by,
\begin{equation} \label{eqn: srpcriterion}
\text{Minimize } \jsrp(\tau) \text{ subject to } \text{FAR}(\tau) \leq \alpha
\end{equation}
where the minimization is over all stopping times $\tau$ such
that $\jsrp(\tau)$ is well-defined. Pollak \cite{pollak85}
established the asymptotic optimality of the
Shiryaev-Roberts-Pollak (SRP) stopping rule for (\ref{eqn:
srpcriterion}).

Another approach to change detection is the Bayesian
formulation of \cite{shiry78, shiry63}. Here the change-point
is modeled as a random variable $\Lambda$ with prior
probability distribution, $\pi_k = \mathsf{P}(\Lambda = k), k =
1,2,\ldots$. The performance measures are the average detection
delay (ADD) and probability of false alarm (PFA) defined by:
\[
\text{ADD} (\tau) = \Expect^\nu [(\tau -\Lambda)^+], \quad \quad \text{PFA}(\tau) = \mathsf{P}^\nu(\tau < \Lambda )
\]
where $\Expect^\nu$ represents the expectation operator and
$\mathsf{P}^\nu$ the probability law when the pre-change and
post-change distributions are $\nu_0$ and $\nu_1$ respectively.
For a given $\alpha \in (0,1)$, the optimization problem under
the Bayesian criterion is:
\begin{equation} \label{eqn: bayescriterion}
\text{Minimize} \text{ ADD}(\tau) \text{ subject to } \text{PFA}(\tau) \leq \alpha
\end{equation}
When the prior distribution on the change-point follows a
geometric distribution, the optimal solution to the above
problem is given by the Shiryaev test \cite{shiry63}.

The robust versions of (\ref{eqn: lordencriterion}), (\ref{eqn:
srpcriterion}) and (\ref{eqn: bayescriterion}) are relevant
when one or both of the distributions $\nu_0$ and $\nu_1$ are
not known exactly, but are known to belong to uncertainty
classes of distributions, $\clP_0, \clP_1 \subset \clP(\clX)$.
The objective is to minimize the worst-case delay amongst all
possible values of the unknown distributions, while satisfying
the false-alarm constraint for all possible values of the
unknown distributions. Thus the robust version of the Lorden
criterion is to identify the stopping rule that solves the
following optimization problem:
\begin{eqnarray}
&\min& \sup_{\nu_0\in \clP_0, \nu_1\in \clP_1}\text{ WDD}(\tau) \label{eqn: lordencriterionrobust}\\
&\text{ s.t. }& \sup_{\nu_0 \in \clP_0}\text{FAR}(\tau) \leq \alpha. \nonumber
\end{eqnarray}
Similarly, the robust version of the SRP criterion is:
\begin{eqnarray}
&\min& \sup_{\nu_0\in \clP_0, \nu_1\in \clP_1}\jsrp(\tau) \label{eqn: srpcriterionrobust}\\
&\text{ s.t. }& \sup_{\nu_0 \in \clP_0}\text{FAR}(\tau) \leq \alpha. \nonumber
\end{eqnarray}
and the robust version of the Bayesian criterion is:
\begin{eqnarray}
&\min& \sup_{\nu_0\in \clP_0, \nu_1\in \clP_1} \text{ ADD}(\tau) \label{eqn: bayescriterionrobust} \\
&\text{ s.t. }& \sup_{\nu_0 \in \clP_0}\text{PFA}(\tau) \leq \alpha \nonumber
\end{eqnarray}

The optimal stopping rule $\tau$ under each of the robust
criteria described above has the following minimax
interpretation. For any other stopping rule $\tau'$ that
guarantees the false alarm constraint for all values of unknown
distributions from the uncertainty classes, there is at least
one pair of distributions such that the delay obtained under
$\tau'$ will be at least as high as the maximum delay obtained
with $\tau$ over all pairs of distributions from the
uncertainty classes. In the rest of this paper we provide
solutions to the robust problems (\ref{eqn:
lordencriterionrobust}), (\ref{eqn: srpcriterionrobust}) and
(\ref{eqn: bayescriterionrobust}) when the uncertainty classes
satisfy some specific conditions.

\section{Robust change detection} \label{sec:robqcd}
\subsection{Least Favorable Distributions}
The solution to the robust problem is simplified greatly if we
can identify least favorable distributions (LFDs) from the
uncertainty classes such that the solution to the robust
problem is given by the solution to the non-robust problem
designed with respect to the LFDs. LFDs were first identified
for a simpler problem - the robust hypothesis testing problem -
by Huber et al. in \cite{huber65} and \cite{hubstrass73}. It
was later shown in \cite{venubasarpoor94} that if the
uncertainty classes satisfy a joint stochastic boundedness
condition, one can identify these LFDs. Before we introduce
this condition, we need the following notation. If $X$ and $X'$
are two real-valued random variables defined on a probability
space $\ofp$ such that,
\[
\Prob(X \geq t) \geq \Prob(X' \geq t), \mbox{ for all } t \in \mathbb{R},
\]
then we say that the random variable $X$ is
\emph{stochastically larger than} \cite{venubasarpoor94} the
random variable $X'$. We denote this relation via the notation
$X \succ X'$. Equivalently if $X \sim \mu$ and $X' \sim \mu'$,
we also denote $\mu \succ \mu'$.

\textit{Definition 1 (Joint Stochastic Boundedness)
\cite{venubasarpoor94}:} Consider the pair $(\clP_0, \clP_1)$
of classes of distributions defined on a measurable space
$(\clX,\clF)$. Let $(\overline \nu_0, \underline\nu_1) \in
\clP_0 \times \clP_1$ be some pair of distributions from this
pair of classes such that $\underline\nu_1$ is absolutely
continuous with respect to $\overline\nu_0$. Let $L^*$ denote
the log-likelihood ratio between $\underline\nu_1$ and
$\overline \nu_0$ defined as the logarithm of the Radon-Nikodym
derivative $\log \frac{d \underline\nu_1}{d \overline \nu_0}$.
Corresponding to each $\nu_j \in \clP_j$, we use $\mu_j$ to
denote the distribution of $L^*(X)$ when $X \sim \nu_j, j =
0,1$. Similarly we use $\overline \mu_0$ (respectively
$\underline \mu_1$) to denote the distribution of $L^*(X)$ when
$X \sim \overline \nu_0$ (respectively $\underline \nu_1$). The
pair $(\clP_0, \clP_1)$ is said to be jointly stochastically
bounded by $(\overline \nu_0, \underline\nu_1)$ if for all
$(\nu_0, \nu_1) \in \clP_0 \times \clP_1$,
\[
\quad \quad \quad \quad \quad \quad \quad \quad \overline \mu_0 \succ \mu_0 \mbox{ and } \mu_1 \succ \underline \mu_1 \quad \quad \quad \quad \quad \quad \quad \blacksquare
\]
Loosely speaking, the LFD from one uncertainty class is the
distribution that is \textit{nearest} to the other uncertainty
class. This notion can be made rigorous in terms of
Kullback-Leibler divergence and other Ali-Silvey distances
between distributions in the uncertainty classes, as shown in
\cite[Corollary 1]{poor80}.

Huber and Strassen \cite{hubstrass73} have established a
procedure to obtain robust solutions to the Neyman-Pearson
hypothesis testing problem provided the uncertainty classes can
be described in terms of 2-alternating capacities. As pointed
out in \cite{venubasarpoor94}, any pair of uncertainty classes
that can be described in terms of 2-alternating capacities also
satisfy the joint stochastic boundedness (JSB) condition (see
\cite[Theorem 4.1]{hubstrass73}). This observation suggests
that we can identify examples of uncertainty classes which
satisfy the joint stochastic boundedness condition using the
results in \cite{hubstrass73}, \cite{venubasarpoor94}, and
\cite{levy08}. These include $\epsilon$-contamination classes,
total variation neighborhoods, Prohorov distance neighborhoods,
band classes, and p-point classes. In general it is difficult
to identify the distributions $\overline \nu_0$ and
$\underline\nu_1$. However, for $\epsilon$-contamination
classes, total variation neighborhoods, and L\'{e}vy metric
neighborhoods, the method suggested in \cite[pp.
241-248]{levy08} can be used to identify these distributions.





We show that under certain assumptions on $\clP_0$  and
$\clP_1$, the pair of distributions $(\overline \nu_0,
\underline \nu_1)$ are LFDs for the robust change detection
problem in (\ref{eqn: lordencriterionrobust}), (\ref{eqn:
srpcriterionrobust}) and (\ref{eqn: bayescriterionrobust}).
Thus the optimal stopping rules designed \textit{assuming}
known pre-change and post-change distributions $\overline
\nu_0$ and $\underline \nu_1$, respectively, are optimal for
the robust problems (\ref{eqn: lordencriterionrobust}),
(\ref{eqn: srpcriterionrobust}) and (\ref{eqn:
bayescriterionrobust}). We use $\Expect_m^*$ to denote the
expectation operator and $\mathsf{P}_m^*$ to denote the
probability law when the change happens at $m$ and the
pre-change and post-change distributions are $\overline \nu_0$
and $\underline \nu_1$, respectively.

We need the following straightforward result. For completeness
we provide a proof in the appendix.

\begin{lemma} \label{lem:incfuncstochord}
Suppose $\{U_i: 1 \leq i \leq n\}$ is a set of mutually
independent random variables, and $\{V_i: 1 \leq i \leq n\}$ is
another set of mutually independent random variables such that
$U_i \succ V_i, 1 \leq i \leq n$. Now let $h: \mathbb{R}^n
\mapsto \mathbb{R}$ be a continuous real-valued function
defined on $\mathbb{R}^n$ that satisfies,
\begin{eqnarray*}
\begin{aligned}
h(x_1, \ldots, x_{i-1}, a&, x_{i+1}, \ldots, x_n)\\
& \geq  \quad h(x_1,\ldots, x_{i-1}, x_i, x_{i+1}, \ldots, x_n),
\end{aligned}
\end{eqnarray*}
for all $x_1^n \in \mathbb{R}^n, a > x_i$, and $i \in
\{1,\ldots,n \}$. Then we have,
\[
h(U_1, U_2, \ldots, U_n) \succ h(V_1, V_2, \ldots, V_n)
\]
\end{lemma}

\subsection{Lorden criterion}\label{sec:Lordencriterion}
When the distributions $\nu_0$ and $\nu_1$ are known, the
solution to (\ref{eqn: lordencriterion}) is given by the CUSUM
test \cite{mous86}. The optimal stopping time is given by,
\begin{eqnarray} \label{eqn:cusum}
\tau_{\scriptscriptstyle \mathrm{C}} = \inf \{n \geq 1: \max_{1\leq k \leq n} \sum_{i=k}^n L^\nu(X_i) \geq \eta \}
\end{eqnarray}
where $L^\nu$ is the log-likelihood ratio between $\nu_1$ and
$\nu_0$, and the threshold $\eta$ is chosen so that,
$\Expect_\infty^\nu (\tau_{\scriptscriptstyle \mathrm{C}}) =
\frac{1}{\alpha}$. The following theorem provides a solution to
the robust Lorden problem when the distributions are unknown.

\begin{theorem} \label{thm:lordenrobust}
Suppose the following conditions hold:
\begin{romannum}
\item The uncertainty classes $\clP_0, \clP_1$ are jointly
    stochastically bounded by $(\overline \nu_0,\underline
    \nu_1)$.
\item All distributions $\nu_0 \in \clP_0$ are absolutely
    continuous with respect to $\overline \nu_0$. i.e.,
\begin{equation}
\nu_0 \ll \overline \nu_0, \quad \nu_0 \in \clP_0. \label{eqn:abscont}
\end{equation}
\item The function $L^*(.)$, representing the log-likelihood
    ratio between $\underline \nu_1$ and $\overline \nu_0$ is
    continuous over the support of $\overline \nu_0$.
\end{romannum}
Then the optimal stopping rule that solves (\ref{eqn:
lordencriterionrobust}) is given by the following CUSUM test:
\begin{eqnarray} \label{eqn:cusumrobust}
\tau_{\scriptscriptstyle \mathrm{C}}^* = \inf \left\{n \geq 1: \max_{1\leq k \leq n} \sum_{i=k}^n L^*(X_i) \geq \eta \right\}
\end{eqnarray}
where the threshold $\eta$ is chosen so that, $\Expect_\infty^*
(\tau_{\scriptscriptstyle \mathrm{C}}^*) = \frac{1}{\alpha}$.
\qed
\end{theorem}
We prove the theorem in the appendix. Two brief remarks are in
order. Firstly, the discussion in \cite[p. 198]{poorhadji09}
suggests that when LFDs exist under our formulation, they also
solve the asymptotic problem, as expected. Secondly, the robust
CUSUM test admits a simple recursive implementation similar to
the ordinary CUSUM test. Clearly,
\begin{equation}\label{eqn:CUSUMrecursion}
S_{n+1} = S_{n}^+ + L^*(X_{n+1}).
\end{equation}
where $S_n = \max_{1\leq k \leq n} \sum_{i=k}^{n} L^*(X_i)$ is
the test statistic appearing in (\ref{eqn:cusumrobust}). Thus
it is easy to compute the test statistic recursively.

\subsubsection{Asymptotic analysis of the robust
CUSUM}\label{sec:CUSUMasym} In general, for any pair of
pre-change and post-change distributions $(\nu_0, \nu_1)$ from
the uncertainty classes, we expect the performance of the
robust CUSUM test to be poorer than that of the optimal CUSUM
test designed with respect to the correct distributions. The
drop in performance can be interpreted as the \textit{cost of
robustness}. Although it is not easy to characterize this cost
in general, some insight can be obtained by performing an
asymptotic analysis in the setting where the false alarm
constraint $\alpha$ goes to zero. Our analysis uses the result
of \cite[Theorem 2]{lorden71} (also see \cite[Theorem
6.16]{poorhadji09}). We use
$\text{WDD}^\nu(\tau_{\scriptscriptstyle \mathrm{C}}^*)$ to
denote the worst-case delay obtained by employing the stopping
rule $\tau_{\scriptscriptstyle \mathrm{C}}^*$ when the
pre-change and post-change distributions are given by $\nu_0$
and $\nu_1$. Similarly, $\text{WDD}^*(\tau_{\scriptscriptstyle
\mathrm{C}}^*)$ is used to denote the same quantity when the
pre-change and post-change distributions are the LFDs.

As mentioned in the remark following Theorem 2 in
\cite{lorden71}, we can interpret the robust CUSUM test as a
repeated one-sided sequential probability ratio test (SPRT)
between $\underline \nu_1$ and $\overline \nu_0$. Let
$\tau_{\scriptscriptstyle \mathrm{SPRT}}$ denote the stopping
rule of the SPRT. We apply \cite[Theorem 2]{lorden71} to
$\tau_{\scriptscriptstyle \mathrm{SPRT}}$ when the true
distributions are the LFDs. It follows that
\[
\Expect_\infty^* (\tau_{\scriptscriptstyle \mathrm{C}}^*) \geq \frac{1}{\alpha}
\]
where $B = \frac{1}{\alpha}$ is used as the upper threshold in
the SPRT given by $\tau_{\scriptscriptstyle \mathrm{SPRT}}$.
From (\ref{eqn:FARordered}), we know that
\[
\Expect_\infty^\nu (\tau_{\scriptscriptstyle \mathrm{C}}^*) \geq \Expect_\infty^* (\tau_{\scriptscriptstyle \mathrm{C}}^*) \geq \frac{1}{\alpha}.
\]
We again apply the theorem to $\tau_{\scriptscriptstyle
\mathrm{SPRT}}$, but with the true distributions given by any
$\nu_0 \in \clP_0$ and $\nu_1 \in \clP_1$. We now have,
\[
\text{WDD}^\nu(\tau_{\scriptscriptstyle \mathrm{C}}^*) \leq \Expect (\tau_{\scriptscriptstyle
\mathrm{SPRT}})
\]
where the expression on the right hand side denotes the
expected stopping time of the SPRT when the observations follow
distribution $\nu_1$. Now, by applying the well-known Wald's
identity \cite{wald47} as suggested in the remark following
\cite[Theorem 2]{lorden71}, we obtain
\[
\Expect (\tau_{\scriptscriptstyle
\mathrm{SPRT}}) = \frac{|\log \alpha|}{I_{\nu_1}}(1 + o(1)), \quad  \mbox{ as } \alpha \to 0
\]
where $o(1) \to 0$ as $\alpha \to 0$ and
\[
I_{\nu_1} = \int L^*(x) d\nu_1(x) = D(\nu_1 \| \overline \nu_0) - D(\nu_1 \| \underline \nu_1).
\]
Thus
\[
\text{WDD}^\nu(\tau_{\scriptscriptstyle \mathrm{C}}^*) \leq \frac{|\log(\alpha)| (1 + o(1))}{D(\nu_1 \| \overline \nu_0) - D(\nu_1 \| \underline \nu_1)}.
\]
It is also known from \cite[Theorem 3]{lorden71} that any
stopping rule $\tau$ that satisfies the false alarm constraint
$\text{FAR}(\tau) \leq \alpha$ must satisfy the lower bound
\[
\text{WDD}^\nu(\tau) \geq \frac{|\log(\alpha)|(1 + o(1))}{D(\nu_1 \| \nu_0)}
\]
and that this lower bound is achieved by the optimal CUSUM test
between $\nu_1$ and $\nu_0$.
%
Thus, the worst-case delay of the robust test is asymptotically
larger by a factor no more than
\[
\frac{D(\nu_1 \| \nu_0)}{D(\nu_1 \| \overline \nu_0) - D(\nu_1 \| \underline \nu_1)}
\]
when compared with the delay incurred by the optimal test. This
factor is thus an upper bound on the asymptotic cost of
robustness.

\subsection{Shiryaev-Roberts-Pollak (SRP) criterion}
\label{sec:srpcriterion}
The SRP stopping rule is asymptotically optimal for (\ref{eqn:
srpcriterion}). Let $R_0^\nu$ be a random variable with
distribution $\psi$ supported on $\Re_+$ and define,
\begin{equation}
R_{n}^\nu = L^\nu(X_{n})(1+R_{n-1}^\nu), \quad n \geq 1. \label{eqn:srpiteration}
\end{equation}
When the distributions $\nu_0$ and $\nu_1$ are known the SRP stopping rule is given by
\begin{eqnarray} \label{eqn:srp}
\tau_{\scriptscriptstyle \mathrm{SRP}}^{\nu, \eta, \psi} = \inf\left\{n \geq 0: R_n^\nu \geq \eta \right\}.
\end{eqnarray}

\noindent \textit{Asymptotic optimality property}: The SRP test
of (\ref{eqn:srp}) is asymptotically optimal for (\ref{eqn:
srpcriterion}) in the following sense \cite{pollak85}: For
every $0 < \alpha < 1$ there exists threshold $\eta$ and
probability measure $\psi_\eta$ such that the stopping rule
$\tau_{\scriptscriptstyle \mathrm{SRP}} :=
\tau_{\scriptscriptstyle \mathrm{SRP}}^{\nu, \eta, \psi_\eta}$
satisfies $\text{FAR}(\tau_{\scriptscriptstyle \mathrm{SRP}}) =
\alpha$ and for any other stopping rule $\tau$ that satisfies
the false alarm constraint $\text{FAR}(\tau) \leq \alpha$, we
have
\begin{equation}
\jsrp(\tau) \geq \jsrp(\tau_{\scriptscriptstyle \mathrm{SRP}})  + o(1) \label{eqn:srpasymproperty}
\end{equation}
where $o(1) \to 0$ as $\alpha \to 0$.

The following theorem identifies a stopping rule that extends
the above asymptotic optimality property to the setting where
the post-change distribution is unknown.
\begin{theorem} \label{thm:srprobust}
Suppose the following conditions hold:
\begin{romannum}
\item The uncertainty class $\clP_0$ is a singleton $\clP_0 =
    \{\nu_0\}$ and the pair $(\clP_0, \clP_1)$ is jointly
    stochastically bounded by $(\nu_0,\underline \nu_1)$.
\item The function $L^*(.)$, representing the log-likelihood
    ratio between $\underline \nu_1$ and $ \nu_0$ is
    continuous over the support of $ \nu_0$.
\end{romannum}
Let $\tsrp^* := \tau_{\scriptscriptstyle \mathrm{SRP}}^{\nu^*,
\eta, \psi_\eta}$ denote the SRP stopping rule defined with
respect to the LFDs $(\nu_0,\underline \nu_1)$, with parameters
$\eta$ and $\psi_\eta$ chosen such that the asymptotic
optimality property of (\ref{eqn:srpasymproperty}) is
satisfied. Then the stopping rule $\tsrp^*$ is also
asymptotically optimal for (\ref{eqn: srpcriterionrobust}) in
the following sense: For every $0 < \alpha < 1$ and for any
stopping rule $\tau$ that satisfies the false alarm constraint
$\text{FAR}(\tau) \leq \alpha$, we have
\begin{equation}
\sup_{\nu^1 \in \clP_1} \jsrp^\nu(\tau) \geq \sup_{\nu^1 \in \clP_1} \jsrp^\nu(\tsrp^*) + o(1) \label{eqn:srprobasymopt}
\end{equation}
where $o(1) \to 0$ as $\alpha \to 0$. \qed
\end{theorem}
The result of (\ref{eqn:srprobasymopt}) can be interpreted as
follows: The difference between the worst-case values of the
delays incurred by the stopping rule $\tsrp^*$ and any other
stopping rule $\tau$ approaches zero as the false alarm
constraint $\alpha$ approaches zero.

Our proof, provided in the appendix, is useful only when
$\clP_0$ is a singleton. It is possible that the asymptotic
optimality result may still hold even for general $\clP_0$,
although the current proof is not applicable. We elaborate on
this further in the discussion in the next section on the
Bayesian criterion, and also in the appendix following the
proof of the theorem.

We also note that in some cases our proof can be adapted to
obtain tests that are exactly optimal for the robust SRP
criterion of (\ref{eqn: srpcriterionrobust}). Polunchenko et
al. \cite{poltar09} study the Shiryaev-Roberts procedure
(SR-\textit{r}) which is identical to the SRP procedure
described earlier, except for the fact that $R_0$ is not random
but fixed at some constant $r$. Theorem 2 of \cite{poltar09}
shows the exact non-asymptotic optimality of the SR-\textit{r}
procedure for detecting a change in distribution from
$\text{Exp}(1)$ to $\text{Exp}(2)$ where $\text{Exp}(\theta)$
refers to an exponential distribution with mean $\theta^{-1}$.
Using that result, the proof of Theorem \ref{thm:srprobust} can
be adapted to obtain the exact robust solution to the
optimization problem in (\ref{eqn: srpcriterionrobust}). In
particular it can be shown that the SR-\textit{r} procedure for
detecting change from $\text{Exp}(1)$ to $\text{Exp}(2)$ given
in \cite[Theorem 2]{poltar09} is also optimal for (\ref{eqn:
srpcriterionrobust}) when $\clP_0 = \{\text{Exp}(1)\}$ and
$\clP_1 = \{\text{Exp}(\theta) : \theta \geq 2\}$.

\subsection{Bayesian criterion}
When the distributions $\nu_0$ and $\nu_1$ are known and the
prior distribution of the change-point is geometric, the
solution to (\ref{eqn: bayescriterion}) is given by the
Shiryaev test \cite{shiry63}. Denoting the parameter of the
geometric distribution by $\rho$, we have,
\[
\pi_k = \rho(1-\rho)^{k-1}, \quad k \geq 1.
\]
The Shiryaev stopping rule is based on comparing the posterior
probability of change to a threshold $\eta'$
\[
\tau_{\scriptscriptstyle \mathrm{S}} = \inf\left\{n \geq 1: \Prob^\nu(\Lambda \leq n| \clF_{n})  \geq \eta' \right\}.
\]
It can be equivalently expressed as,
\begin{eqnarray} \label{eqn:shiryaev}
\tau_{\scriptscriptstyle \mathrm{S}} = \inf\left\{n \geq 1: \log ( \sum_{k = 1}^n \pi_k \exp (\sum_{i=k}^n L^\nu(X_i) )  )  \geq \eta \right\}
\end{eqnarray}
where the threshold $\eta$ is chosen such that
$\text{PFA}(\tau_{\scriptscriptstyle \mathrm{S}}) =
\mathsf{P}^\nu(\tau_{\scriptscriptstyle \mathrm{S}} < \Lambda )
= \alpha$. The following theorem, proved in the appendix,
identifies a solution to the robust Shiryaev problem (\ref{eqn:
bayescriterionrobust}).
\begin{theorem} \label{thm:bayesrobust}
Suppose the following conditions hold:
\begin{romannum}
\item The uncertainty class $\clP_0$ is a singleton $\clP_0 =
    \{\nu_0\}$ and the pair $(\clP_0, \clP_1)$ is jointly
    stochastically bounded by $(\nu_0,\underline \nu_1)$.
\item The prior distribution of the change-point is a geometric
    distribution.
\item The function $L^*(.)$, representing the log-likelihood
    ratio between $\underline \nu_1$ and $\nu_0$, is
    continuous over the support of $\nu_0$.
\end{romannum}
Then the optimal stopping rule that solves (\ref{eqn:
bayescriterionrobust}) is given by the following Shiryaev test:
\begin{eqnarray} \label{eqn:shiryaevrobust}
\tau_{\scriptscriptstyle \mathrm{S}}^* = \inf\left\{n \geq 1: \log ( \sum_{k = 1}^n \pi_k \exp( \sum_{i=k}^n L^*(X_i) )  )  \geq \eta \right\}
\end{eqnarray}
where the threshold $\eta$ is chosen so that
$\mathsf{P}^*(\tau_{\scriptscriptstyle \mathrm{S}}^* < \Lambda
) = \alpha$. \qed
\end{theorem}

We note that our results under the Bayesian and SRP criteria
are applicable only when the pre-change distribution is known
exactly and hence these results are weaker than our result
under the Lorden criterion. Suppose $\clP_0$ is not a singleton
and $(\clP_0, \clP_1)$ is jointly stochastically bounded by
$(\overline \nu_0,\underline \nu_1)$. In this case, the
stopping rule $\tau_{\scriptscriptstyle \mathrm{S}}^*$ defined
with respect to $(\overline \nu_0, \underline \nu_1)$ is not
optimal for the robust Bayesian criterion (\ref{eqn:
bayescriterionrobust}). In particular, when the pre-change
distribution is $\nu_0 \neq \overline \nu_0$ and the
post-change distribution is $\nu_1 = \underline \nu_1$, it can
be shown that the average detection delay
$\text{ADD}^\nu(\tau_{\scriptscriptstyle \mathrm{S}}^*)$ of the
stopping rule $\tau_{\scriptscriptstyle \mathrm{S}}^*$ is in
general higher than the average detection delay
$\text{ADD}^*(\tau_{\scriptscriptstyle \mathrm{S}}^*)$ when the
pre-change and post-change distributions are $(\overline
\nu_0,\underline \nu_1)$. This is because the likelihood ratios
of the pre-change observations appearing in
(\ref{eqn:shiryaevrobust}) are stochastically larger under
$\overline \nu_0$ than under $\nu_0$. This leads to a stopping
time that is stochastically smaller under $(\overline
\nu_0,\underline \nu_1)$ than under $(\nu_0,\underline \nu_1)$.
Hence there is no reason to believe that
$\tau_{\scriptscriptstyle \mathrm{S}}^*$ solves the robust
problem (\ref{eqn: bayescriterionrobust}).

Even in the case of the SRP criterion studied in Section
\ref{sec:srpcriterion}, our robust result holds only when
$\clP_0$ is a singleton and the JSB condition holds. However,
unlike in the Bayesian case, we do not have a simple
explanation for why the result cannot be extended to the
setting where the pre-change distribution is not known exactly.
It is possible that for some specific choices of the
uncertainty classes, the stopping rule designed with respect to
$(\overline \nu_0,\underline \nu_1)$ may be asymptotically
optimal for the robust problem of (\ref{eqn:
srpcriterionrobust}), although we do not expect this to be true
in general.

However, such a problem does not arise for the robust CUSUM
test we studied in Section \ref{sec:Lordencriterion}, since the
worst-case detection delay
$\text{WDD}^\nu(\tau_{\scriptscriptstyle \mathrm{C}}^*)$ of the
robust CUSUM depends only on the support of the pre-change
distribution when post-change distribution is kept fixed at
$\nu_1 = \underline \nu_1$.

\medskip

\noindent \textit{Comparison with robust sequential detection}
\quad It is interesting to compare our results with some known
results on robust sequential detection. We have shown that
ptovided the JSB condition and other regularity conditions
hold, change detection tests designed with respect to the LFDs
exactly solve the minimax robust change detection problem under
the Lorden and Bayesian criteria. However, the known minimax
optimality results in robust sequential detection are all for
the asymptotic settings - as error probabilities go to zero
\cite{huber65} or as the size of the uncertainty classes
diminishes \cite{qua85}. Huber \cite{huber65} showed that an
exact minimax result does not hold for the robust sequential
detection problem in general. He provided examples where the
expected stopping times of the SPRT designed with respect to
the LFDs are not least favorable under the LFDs. This is
similar to the reason why the robust Shiryaev test is not
optimal for the Bayesian problem when $\clP_0$ is not a
singleton as explained above.

\begin{figure}
\centering
\includegraphics[width=0.7\textwidth]{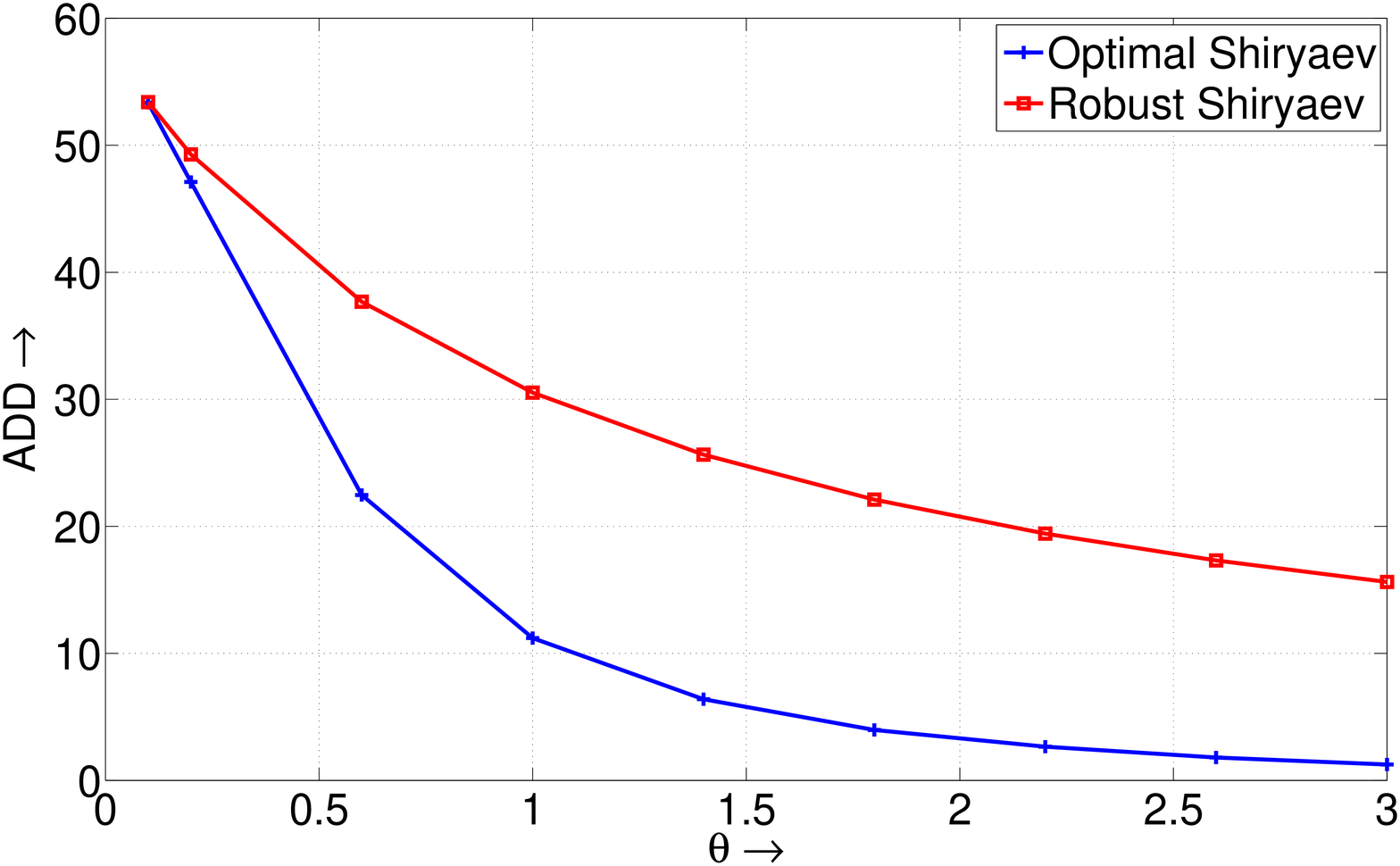} 
 \caption{Comparison of robust and non-robust Shiryaev tests for $\alpha = 0.001$ for the Gaussian mean shift example.}  \label{figGaussBayes}
\end{figure}

%
\section{ Some examples and simulation results} \label{sec:example}
\subsection{Gaussian mean shift}
Here we consider a simple example to illustrate the results.
Assume $\nu_0$ is known to be a standard Gaussian distribution
with mean zero and unit variance, so that $\clP_0$ is a
singleton. Let $\clP_1$ be the collection of Gaussian
distributions with means from the interval $[0.1,3]$ and unit
variance.
\begin{eqnarray}
\clP_0 &=& \{ \clN(0,1)\} \nonumber \\
\clP_1 &=& \{\clN(\theta,1): \theta \in [0.1,3] \} \label{eqn:Gaussmeanshift}
\end{eqnarray}
It is easily verified that $(\clP_0,\clP_1)$ is jointly
stochastically bounded by $(\overline \nu_0,\underline \nu_1)$
given by
\[
\overline \nu_0 \sim \clN(0,1), \qquad \underline \nu_1 \sim \clN(0.1,1).
\]

\subsubsection{Bayesian criterion}
We simulated the Bayesian and robust Bayesian change detection
tests for this problem assuming a geometric prior distribution
for the change-point with parameter $0.1$ and a false alarm
constraint of $\alpha = 0.001$. From the performance curves
plotted in Figure \ref{figGaussBayes}, we can see that the
robust Shiryaev test gives the same average detection delay
(ADD) as the optimal Shiryaev test at $\underline \nu_1$ which
corresponds to $\theta = 0.1$ in the figure. This is expected
since the robust test is identical to the optimal test at
$\underline \nu_1$. For all other values of $\nu_1 \in \clP_1$,
the performance of the robust test is strictly better than the
performance at $\underline \nu_1$ and hence this test is indeed
minimax optimal. We also see in Figure \ref{figGaussBayes} that
the average delays obtained with the robust test are much
higher than those obtained with the optimal test, especially at
high values of the mean $\theta$. The probability of false
alarm and average detection were estimated  via Monte-Carlo
simulations with a standard deviation of $0.1\%$ for the
estimates.

%

\subsubsection{Lorden criterion and comparison with GLR test}
Under the Lorden criterion, we compared the performances of
three tests - the optimal CUSUM test with known $\theta$, the
robust CUSUM test designed with respect to the LFDs, and the
CUSUM test based on the Generalized Likelihood Ratio (GLR test)
suggested in \cite{lorden71}. The stopping time under the GLR
test is given by
\begin{equation}
\tau_{\scriptscriptstyle \mathrm{GLR}} = \inf \{n \geq 1:  \max_{1\leq k \leq n} \sup_{\nu_1 \in \clP_1} \sum_{i=k}^n L^\nu(X_i) \geq \eta \} \label{eqn:GLRT}
\end{equation}
where $\eta$ is chosen so that the false alarm constraint is
met with equality. The GLR test does not require knowledge of
$\theta$ but still achieves the same asymptotic performance as
the optimal CUSUM test with known $\theta$ when the false alarm
constraint goes to zero for some choices of the uncertainty
classes including the example considered above.

\begin{figure}
\centering
\includegraphics[width=0.7\textwidth]{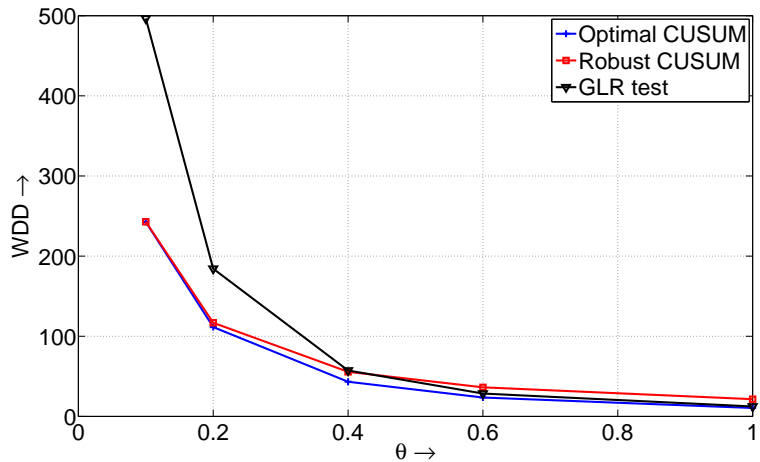}
\caption{Comparison of various tests for false alarm rate of
$\alpha = 0.001$ for the Gaussian mean shift example.} \label{figGaussLordenGLRT}
\end{figure}

Figure \ref{figGaussLordenGLRT} and Table \ref{tbl:Lordensims}
shows estimates of the worst-case detection delay (WDD)
obtained under the these tests designed for a false alarm
constraint of $\alpha = 0.001$, for various values of $\theta$.
These values are estimated using Monte-Carlo simulations. The
delay values have a standard deviation lower than $1\%$ and the
false alarm value has a standard deviation lower than $3\%$.


From the performance curves in Figure \ref{figGaussLordenGLRT}
and the values in Table \ref{tbl:Lordensims} we see that the
GLR test gives better performance than our robust solution at
higher values of $\theta$, and is close to optimal at these
high values of $\theta$. However, the robust test gives much
better performance than the GLR test at the low values of
$\theta$. This is expected since the robust solution is minimax
optimal and hence is expected to perform better at the
unfavorable values of $\theta$.

An important difference between the two solutions is that
although the robust CUSUM test based on the LFDs admits a
simple recursive implementation like we described in
(\ref{eqn:CUSUMrecursion}), the GLR test is in general very
complex to implement. This is because the supremum in
(\ref{eqn:GLRT}) may be achieved at different values of $\nu_1$
for different $n$. Furthermore, the optimization in
(\ref{eqn:GLRT}) may not be easy to solve for general
uncertainty classes - particularly non-parametric classes like
the $\epsilon$-uncertainty classes considered next.

\begin{table}
\begin{center}
\caption{Delays obtained using various tests under the Lorden
criterion for a false alarm rate of $\alpha = 0.001$.}
\label{tbl:Lordensims}
\begin{tabular}{|c|c|c|c|}
\hline $\theta$ & Optimal CUSUM & Robust CUSUM & GLR test\\
\hline 0.1 &242.7 & 242.7 & 496\\
0.2 &111.5 & 116.8 & 184\\
0.4  &43.2 & 55.6 & 57.2\\
0.6   &23.5 & 36.3 & 28.6\\
1.0 & 10.5& 21.5 & 12.35\\ \hline
\end{tabular}
\end{center}
\end{table}

%
%

\subsection{$\epsilon$-contamination classes}
We now discuss an example in which the uncertainty class
$\clP_0$ is no longer a singleton. For some scalar $\epsilon
\in (0,1)$, consider the following $\epsilon$-contamination
classes:
\begin{eqnarray}
\clP_0 = \{\nu_0: \nu_0 = (1-\epsilon) {\clN(0,1)}  + \epsilon H_0, \quad H_0 \in \clP(\Re)\} \\
\clP_1 = \{\nu_1: \nu_1 = (1-\epsilon) {\clN(1,1)}  + \epsilon H_1, \quad H_1 \in \clP(\Re)\}
\end{eqnarray}
where $\clP(\Re)$ is the collection of all probability measures
on $\Re$ and $\clN(\mu,\sigma)$ denotes the probability measure
corresponding to a Gaussian random variable with mean $\mu$ and
variance $\sigma^2$. In other words, the distributions in
uncertainty class $\clP_i$ are mixtures of a Gaussian
distribution with mean $i$ and unit variance, and an arbitrary
probability distribution on $\Re$ with weights given by $1-
\epsilon$ and $\epsilon$ respectively.

Following the method outlined in \cite{huber65}, we identified
LFDs for these uncertainty classes and evaluated the
performance of the robust test. Let $p_i$ denote the density
function of a $\clN(\mu,1)$ random variable and let $q_i$
denote the density function of the least favorable distribution
from $\clP_i$. It is established in \cite{huber65} that the
densities of the LFDs have the following structure,
\begin{eqnarray}\label{eqn:LFDpdf}
q_0(x) &=& \left\{ \begin{tabular}{cc} $(1-\epsilon) p_0(x)  $ &if $L(x) \leq b$\\
$\frac{1-\epsilon}{b} p_1(x)$ &if $L(x) > b$
\end{tabular} \right.\\
q_1(x) &=& \left\{ \begin{tabular}{cc} $(1-\epsilon) p_1(x)  $ &if $L(x) > a$\\
$a(1-\epsilon) p_0(x)$ &if $L(x) \leq a$
\end{tabular} \right.
\end{eqnarray}
where $L(x) = \frac{p_1(x)}{p_0(x)}$. The scalars $a$ and $b$
are identified by the following relation:
\begin{eqnarray*}
(1-\epsilon) \int_{ \{x: L(x) \leq b \}} p_0(x) dx + \frac{1-\epsilon}{b} \int_{ \{x: L(x) > b \}} p_1(x) dx &=& 1\\
(1-\epsilon) \int_{ \{x: L(x) > a \}} p_1(x) dx + a(1-\epsilon) \int_{ \{x: L(x) \leq a \}} p_0(x) dx &=& 1.
\end{eqnarray*}

In order to compare the performance of the robust test with
that of the optimal test we chose the following distributions
for $H_0$ and $H_1$:
\[
H_0 = \clN(0,\sigma_0), \sigma_0 \in [0.1, 10] \qquad \qquad H_1 = \clN(1,\sigma_1), \sigma_1 \in [0.1, 10].
\]
Table \ref{tbl:epssimsig0fixd} shows the values of the
worst-case delay (WDD) obtained when $\sigma_0$ is kept fixed
at $\sigma_0=1$ and $\sigma_1$ is varied. Shown are the results
obtained using the robust CUSUM test as well as the optimal
CUSUM test for $\epsilon = 0.05$ and for $\epsilon = 0.005$. We
notice that the difference in performance between the robust
test and the optimal test is larger for larger values of
$\epsilon$. This matches the intuition that the cost of
robustness would be higher for a larger uncertainty class of
distributions. The delay values and false alarm rates were
estimated to have standard deviations lower than $0.1\%$ and
$1\%$ respectively.

%
%
%

Table \ref{tbl:epssimsig1fixd} shows the values of worst-case
delay obtained under the optimal CUSUM tests when $\sigma_1$ is
kept fixed at $\sigma_1=1$ and $\sigma_0$ is varied. The delay
values and false alarm rates were estimated to have standard
deviations lower than $0.1\%$ and $1\%$ respectively. We have
not included the delays obtained under the robust test, since
the delay of the robust test is invariant with $\sigma_0$. The
delay obtained under the robust test for $\epsilon = 0.05$ and
$\epsilon = 0.005$ are respectively $15.09$ and $11.27$ as
shown in the third row of Table \ref{tbl:epssimsig0fixd}
corresponding to $\sigma_1 = 1$.

\begin{table}
\begin{center}
\caption{Delays obtained using various tests under the Lorden
criterion for $\epsilon$-uncertainty classes with $\alpha =
0.001$ and $\sigma_0 = 1$. } \label{tbl:epssimsig0fixd}
\begin{tabular}{|c|c|c|c|c|}
\hline &\multicolumn{2}{c|}{$\epsilon = 0.05$} &\multicolumn{2}{c|}{$\epsilon = 0.005$}\\
\cline{2-5}
$\sigma_1$ & Robust CUSUM & Optimal CUSUM & Robust CUSUM& Optimal CUSUM\\
\hline 0.1 & 14.77 & 9.17  & 11.27& 10.38\\
0.5 & 14.86&9.12  & 11.27 &10.39\\
1  & 15.09& 9.08 & 11.27 &10.35\\
5   &15.52 & 8.78 & 11.29 & 10.33\\
10 & 15.59& 8.65 & 11.29 & 10.34\\ \hline
\end{tabular}
\end{center}
\end{table}

\begin{table}
\begin{center}
\caption{Delays obtained using the optimal CUSUM test for
$\epsilon$-uncertainty classes with $\alpha = 0.001$ and
$\sigma_1 = 1$.} \label{tbl:epssimsig1fixd}
\begin{tabular}{|c|c|c|}
\hline $\sigma_0$ & Optimal CUSUM for $\epsilon = 0.05$ & Optimal CUSUM for $\epsilon = 0.005$ \\
\hline 0.1 & 10.56 & 10.55\\
0.5 & 10.50&10.52  \\
1  & 10.44& 10.56\\
5   &10.02 &10.58\\
10 & 9.85& 10.59\\ \hline
\end{tabular}
\end{center}
\end{table}

%


\section{Conclusion} \label{sec:conclusion}
We have shown that for uncertainty classes that satisfy some
specific conditions, the optimal change detectors designed for
the least favorable distributions are optimal in a minimax
sense. This is shown for the Lorden criterion, the
Shiryaev-Roberts-Pollak criterion, and Shiryaev's Bayesian
criterion. However, robustness comes at a potential cost. The
optimal stopping rule designed for the LFDs may perform quite
sub-optimally for other distributions from the uncertainty
class when compared with the optimal performance that can be
obtained in the case where these distributions are known
exactly. Using an asymptotic analysis, we have also obtained an
analytic upper bound on this cost of robustness for the robust
solution under the Lorden criterion. Nevertheless for some
parameter ranges our robust test obtains significant
performance improvement over the CUSUM test designed for the
Generalized Likelihood Ratio statistic, which is a benchmark
for the composite quickest change detection problem. Our robust
solution also has the added advantage that it can be
implemented in a simple recursive manner, while the GLR test
does not admit a recursive solution in general, and may require
the solution to a complex non-convex optimization problem at
every time instant.



\appendix
\section{Appendix}

\subsection{Proof of Lemma \ref{lem:incfuncstochord}}

We prove this claim by induction. For $n = 1$, the claim holds
because if $h : \mathbb{R} \mapsto \mathbb{R}$ is a
non-decreasing continuous function we have,
\begin{eqnarray*}
\Prob (h(U_1) \geq t) &=& \Prob (U_1 \geq \sup \{x : h(x) < t \} \\
&\geq& \Prob (V_1 \geq \sup \{x : h(x) < t \} \\
&=& \Prob (h(V_1) \geq t).
\end{eqnarray*}

Assume the claim is true for $n = N$ and now consider $n =
N+1$. For any fixed $x_1^N \in \mathbb{R}^N$, since the
function $h$ is non-decreasing in each of its components, it
follows by the proof for $n=1$ that,
\begin{equation}
h(x_1, x_2, \ldots, x_N, U_{N+1})  \succ h(x_1, x_2, \ldots, x_N, V_{N+1}). \label{eqn:gordered}
\end{equation}
We further have,
\begin{eqnarray}
\lefteqn{\Prob (h(U_1, U_2, \ldots, U_{N+1}) \geq t)} \nonumber \\
&=& \int f_{U_1^N}(x_1^N) \Prob (h(x_1, x_2, \ldots, x_N, U_{N+1}) \geq t ) dx_1^N \nonumber \\
&\geq& \int f_{U_1^N}(x_1^N) \Prob (h(x_1, x_2, \ldots, x_N, V_{N+1}) \geq t ) dx_1^N\label{eqn:gorderused}\\
&=& \Prob (h(\tilde U_1, \tilde U_2, \ldots, \tilde U_N, V_{N+1}) \geq t) \label{eqn:tildex}\\
&=& \int f_{V_{N+1}}(y) \Prob (h(\tilde U_1, \tilde U_2, \ldots, \tilde U_N, y) \geq t )dy \nonumber \\
&\geq& \int f_{V_{N+1}}(y) \Prob (h(V_1, V_2, \ldots, V_N, y) \geq t )dy \label{eqn:inductionhypused}\\
&=& \Prob (h(V_1, V_2, \ldots, V_{N+1}) \geq t). \nonumber
\end{eqnarray}
where (\ref{eqn:gorderused}) is obtained via
(\ref{eqn:gordered}). The variables $\tilde U_i$ appearing in
(\ref{eqn:tildex}) are random variables with exact same
statistics as $U_i$ and independent of $V_i$'s. The inequality
of (\ref{eqn:inductionhypused}) is obtained by using the
induction hypothesis for $n = N$. Thus we have shown that,
\[
h(U_1, U_2, \ldots, U_{N+1}) \succ h(V_1, V_2, \ldots, V_{N+1})
\]
which proves the lemma by the principle of mathematical
induction. \qed

\subsection{Proof of Theorem \ref{thm:lordenrobust}}
\begin{proof}
Suppose $\clP_0$ and $\clP_1$ satisfy the conditions of the
theorem. Since the CUSUM test is optimal for known
distributions, it is clear that the test given in
(\ref{eqn:cusumrobust}) is optimal when the pre- and
post-change distributions are $\overline \nu_0$ and $\underline
\nu_1$, respectively. Hence, it suffices to show that the
values of $\text{WDD}(\tau_{\scriptscriptstyle \mathrm{C}}^*)$
and $\text{FAR}(\tau_{\scriptscriptstyle \mathrm{C}}^*)$
obtained under any $\nu_0 \in \clP_0$ and any $\nu_1 \in
\clP_1$, are no higher than their respective values when the
pre- and post-change distributions are $\overline \nu_0$ and
$\underline \nu_1$. We use $Y_i^*$ to denote the random
variable $L^*(X_i)$ when the pre-change and post-change
distributions of the observations from the sequence $\{X_i: i =
1,2, \ldots\}$ are $\overline \nu_0$ and $\underline \nu_1$,
respectively, and $Y_i^\nu$ to denote the random variable
$L^*(X_i)$ when the pre- and post-change distributions are
$\nu_0$ and $\nu_1$, respectively. We first prove the theorem
for a special case.

\medskip

\noindent \emph{Case 1:} $\clP_0$ is a singleton given by
$\clP_0 = \{ \nu_0\}$.

Clearly, in this case $\overline \nu_0 = \nu_0$ and
(\ref{eqn:abscont}) is met trivially. Furthermore, in this
case, the false alarm constraint is also met trivially since
the false alarm rate obtained by using the stopping rule
$\tau_{\scriptscriptstyle \mathrm{C}}^*$ is independent of the
true value of the post-change distribution. Fix the
change-point to be $\lambda$. Now, to complete the proof for
the scenario where $\clP_0$ is a singleton, we will show that
for all $\lambda \geq 1$,
\begin{eqnarray}
\Expect_\lambda^*[ (\tau_{\scriptscriptstyle \mathrm{C}}^* - \lambda + 1)^+ | \clF_{\lambda-1} ] \succ 
\Expect_\lambda^\nu[ (\tau_{\scriptscriptstyle \mathrm{C}}^* - \lambda + 1)^+ | \clF_{\lambda-1} ] \label{eqn:delayordered}
\end{eqnarray}
which will establish that the value of
$\text{WDD}(\tau_{\scriptscriptstyle \mathrm{C}}^*)$, obtained
under any $\nu_1 \in \clP_1$, is no higher than the value when
the true post-change distribution is $\underline \nu_1$.

Since we now have $\overline \nu_0 = \nu_0$, both $Y_i^*$ and
$Y_i^\nu$ have the same distributions for $i < \lambda$ and
hence we assume without loss of generality that for all $i <
\lambda$, $Y_i^* = Y_i^\nu$ with probability one. Under this
assumption, we will show that for all integers $N \geq 0$, the
following relation holds with probability one,
\begin{eqnarray}
\begin{aligned}
\mathsf{P}^*_\lambda ((\tau_{\scriptscriptstyle \mathrm{C}}^* - \lambda &+ 1)^+ \leq N | \clF_{\lambda-1}) \\
& \leq \quad \mathsf{P}^\nu_\lambda ((\tau_{\scriptscriptstyle \mathrm{C}}^* - \lambda + 1)^+ \leq N | \clF_{\lambda-1}), \label{eqn:delayordered2}
\end{aligned}
\end{eqnarray}
which will then establish (\ref{eqn:delayordered}).
Since $\tau_{\scriptscriptstyle \mathrm{C}}^*$ is a stopping
time, the event $\{(\tau_{\scriptscriptstyle \mathrm{C}}^* -
\lambda + 1)^+ \leq 0\}$ is $\clF_{\lambda-1}$-measurable.
Hence, with probability one,
(\ref{eqn:delayordered2}) holds with equality for $N=0$. Now it
suffices to verify (\ref{eqn:delayordered2}) for $N \geq 1$. We
know by the stochastic ordering condition on $\clP_1$ that,
\begin{equation}
Y_i^\nu \succ Y_i^*, \mbox{ for all } i \geq \lambda \label{eqn:LLRstochorder}
\end{equation}
Now we have the following equivalence between two events:
\begin{eqnarray*}
\{\tau_{\scriptscriptstyle \mathrm{C}}^* \leq N \} &=& \left\{ \max_{1\leq n \leq N}  \max_{1\leq k \leq n} \sum_{i=k}^n L^*(X_i) \geq \eta  \right\} \\
&=& \left\{ \max_{1\leq k \leq n \leq N}  \sum_{i=k}^n L^*(X_i) \geq \eta  \right\}.
\end{eqnarray*}
It is easy to see that the function,
\[
f(x_1, \ldots, x_N) \triangleq \max_{1\leq k \leq n \leq N}  \sum_{i=k}^n x_i
\]
is continuous and non-decreasing in each of its components as
required by Lemma \ref{lem:incfuncstochord}. Hence for $N \geq
1$, the following hold with probability one:
\begin{eqnarray*}
\begin{aligned}
\mathsf{P}^*_\lambda ((\tau_{\scriptscriptstyle \mathrm{C}}^* - \lambda + 1&)^+ \leq N | \clF_{\lambda-1}) \\
&= \quad \mathsf{P}^*_\lambda (\tau_{\scriptscriptstyle \mathrm{C}}^*  \leq N + \lambda - 1 | \clF_{\lambda-1})\\
&= \quad \mathsf{P}_\lambda (f(Y_1^*, \ldots, Y_{N + \lambda - 1}^*) \geq \eta| \clF_{\lambda-1}) \\
&\leq \quad \mathsf{P}_\lambda (f(Y_1^\nu, \ldots, Y_{N + \lambda - 1}^\nu) \geq \eta| \clF_{\lambda-1})\\
&= \quad \mathsf{P}^\nu_\lambda (\tau_{\scriptscriptstyle \mathrm{C}}^* \leq N | \clF_{\lambda-1}) \\
&= \quad \mathsf{P}^\nu_\lambda ((\tau_{\scriptscriptstyle \mathrm{C}}^* - \lambda + 1)^+ \leq N | \clF_{\lambda-1})
\end{aligned}
\end{eqnarray*}
where the inequality follows from Lemma
\ref{lem:incfuncstochord} and (\ref{eqn:LLRstochorder}), using
the fact that $f$ is a non-decreasing function with respect to
its last $N$ arguments and the fact that $Y_i^\nu = Y^*_i$ for
$i < \lambda$. Thus, for all integers $N \geq 0$,
(\ref{eqn:delayordered2}) holds with probability one and hence
(\ref{eqn:delayordered}) is satisfied. This proves the result
for the case where $\clP_0$ is a singleton.

\medskip
\noindent \emph{Case 2:} $\clP_0$ is any class of distributions
satisfying (\ref{eqn:abscont}).

Suppose that the change does not occur. Then we know by the
stochastic ordering condition on $\clP_0$ that, $Y_i^* \succ
Y_i^\nu$ for all $i$. It follows by Lemma
\ref{lem:incfuncstochord} that,
\begin{eqnarray*}
\mathsf{P}^*_\infty (\tau_{\scriptscriptstyle \mathrm{C}}^* \leq N ) &=& \mathsf{P}_\infty (f(Y_1^*, \ldots, Y_N^*) \geq \eta) \\
&\geq& \mathsf{P}_\infty (f(Y_1^\nu, \ldots, Y_N^\nu) \geq \eta)\\
&=& \mathsf{P}^\nu_\infty (\tau_{\scriptscriptstyle \mathrm{C}}^* \leq N )
\end{eqnarray*}
Since the above relation holds for all $N \geq 1$, we have
\begin{equation}
\Expect_\infty^\nu (\tau_{\scriptscriptstyle
\mathrm{C}}^*) \geq \Expect_\infty^* (\tau_{\scriptscriptstyle
\mathrm{C}}^*) = \frac{1}{\alpha} \label{eqn:FARordered}
\end{equation}
and hence the value of $\text{FAR}(\tau_{\scriptscriptstyle
\mathrm{C}}^*)$ is no higher than $\alpha$ for all values of
$\nu_0 \in \clP_0$ and $\nu_1 \in \clP_1$.

Now suppose the change-point is fixed at $\lambda$. A useful
observation is that for any given stopping rule $\tau$ and
fixed post-change distribution $\nu_1$, the random variable
$\Expect_\lambda^{\nu_0, \nu_1}[ (\tau - \lambda + 1)^+ |
\clF_{\lambda-1} ]$ is a fixed deterministic function of the
random observations $(X_1, \ldots, X_{\lambda-1})$,
irrespective of the distribution $\nu_0$. Thus the essential
supremum of this random variable depends only on the support of
$\nu_0$. Applying this observation to the stopping rule
$\tau_{\scriptscriptstyle \mathrm{C}}^*$, and using the
relation (\ref{eqn:abscont}), we have for all $\nu_0 \in
\clP_0, \nu_1 \in \clP_1$,
\begin{eqnarray*}
\begin{aligned}
\esssup \Expect_\lambda^{\nu_0, \nu_1}[&(\tau_{\scriptscriptstyle \mathrm{C}}^* - \lambda + 1)^+ | \clF_{\lambda-1}] \\
& \leq \quad \esssup \Expect_\lambda^{\overline \nu_0, \nu_1}[ (\tau_{\scriptscriptstyle \mathrm{C}}^* - \lambda + 1)^+ | \clF_{\lambda-1}].
\end{aligned}
\end{eqnarray*}
We also know from \emph{Case 1} above that for all $\nu_1 \in
\clP_1$,
\begin{eqnarray*}
\begin{aligned}
\esssup \Expect_\lambda^{\overline \nu_0, \nu_1}[&(\tau_{\scriptscriptstyle \mathrm{C}}^* - \lambda + 1)^+ | \clF_{\lambda-1}] \\
& \leq \quad \esssup \Expect_\lambda^*[ (\tau_{\scriptscriptstyle \mathrm{C}}^* - \lambda + 1)^+ | \clF_{\lambda-1}].
\end{aligned}
\end{eqnarray*}
Taking the supremum over $\lambda \geq 1$, it follows from the
above two relations that the value of
$\text{WDD}(\tau_{\scriptscriptstyle \mathrm{C}}^*)$ under any
pair of distributions $(\nu_0,\nu_1) \in \clP_0 \times \clP_1$
is no larger than that under $(\overline \nu_0, \underline
\nu_1)$. Thus $\tau_{\scriptscriptstyle \mathrm{C}}^*$ solves
the robust problem (\ref{eqn: lordencriterionrobust}).
\end{proof}

\subsection{Proof of Theorem \ref{thm:srprobust}}
\begin{proof}
Let $\tsrp^* := \tau_{\scriptscriptstyle \mathrm{SRP}}^{\nu^*,
\eta, \psi_\eta}$ denote the SRP stopping rule defined with
respect to the LFDs $(\nu_0,\underline \nu_1)$ satisfying the
asymptotic optimality property of (\ref{eqn:srpasymproperty})
as mentioned in the statement of the theorem. It is easy to see
that for any integers $\lambda \geq 1$ and $N \geq 1$, we have
\begin{eqnarray*}
\begin{aligned}
\Prob_\lambda^\nu(\tsrp^* -\lambda \leq N &| \tsrp^* \geq \lambda, R_0^* = r ) \\
&= \quad \frac{\Prob_\lambda^\nu(\{\tsrp^* -\lambda \leq N\} \cap \{ \tsrp^* \geq \lambda\} | R_0^* = r )}{\Prob_\lambda^\nu( \tsrp^* \geq \lambda | R_0^* = r)}
\end{aligned}
\end{eqnarray*}
where $R_0^*$ denotes the random variable with distribution
$\psi_\eta$ used for initializing the iteration in
(\ref{eqn:srpiteration}). We follow the same steps as in the
proof of Theorem \ref{thm:lordenrobust}. Let $Y^\nu_i$ denote
the random variable $L^*(X_i)$ when the pre-change and
post-change distributions are $\nu_0$ and $\nu_1$ respectively.
Since $\tsrp^*$ is a stopping time the event $\{\tsrp^* \geq
\lambda\}$ is measurable with respect to the pre-change
observations and hence we can represent this event as,
\[
\{\tsrp^* \geq \lambda\} = \{(Y^\nu_1, Y^\nu_2, \ldots, Y^\nu_{\lambda-1}) \in T \}
\]
where $T$ is the set of pre-change trajectories corresponding
to the event $\{\tsrp^* \geq \lambda\}$. Hence, for any $r \in
\Re_+$, we can express the conditional probability as
\begin{eqnarray} \label{eqn:srpint}
\begin{aligned}
\Prob_\lambda^\nu&(\tsrp^* -\lambda \leq N | \tsrp^* \geq \lambda, R_0^* = r ) \\
&= \quad \frac{\int_T \Prob_\lambda( f_t(Y^\nu_\lambda, Y^\nu_{\lambda+1}, \ldots, Y^\nu_{\lambda+N}) \geq g(t,N,\eta) | R_0^* = r)  \Prob_\lambda( (Y^\nu_1, Y^\nu_2, \ldots, Y^\nu_{\lambda-1}) \in dt | R_0^* = r)  }{\int_T \Prob_\lambda( (Y^\nu_1, Y^\nu_{2}, \ldots, Y^\nu_{\lambda-1}) \in dt | R_0^*=r)  }
\end{aligned}
\end{eqnarray}
such that for all $t \in T$, the function $f_t:
\Re^{N-\lambda+1} \mapsto \Re$ satisfies the requirements of
Lemma \ref{lem:incfuncstochord}, and $g(t,N,\eta)$ is some
real-valued function. The exact form of function $f_t$ can be
obtained from the iterations of (\ref{eqn:srpiteration}) used
to define the SRP stopping rule of (\ref{eqn:srp}). We note
that in (\ref{eqn:srpint}) the post-change distribution $\nu_1$
affects only the first term under the integral in the
numerator. Thus, it follows by applying Lemma
\ref{lem:incfuncstochord} that
\begin{eqnarray}\label{eqn:srpstochorder}
\begin{aligned}
\Prob_\lambda^*(\tsrp^* -\lambda \leq N &| \tsrp^* \geq \lambda, R_0^* = r ) \\
&\leq \Prob_\lambda^\nu(\tsrp^* -\lambda \leq N | \tsrp^* \geq \lambda, R_0^* = r )
\end{aligned}
\end{eqnarray}
for all $\nu_1 \in \clP_1$. Hence it further follows that,
\begin{eqnarray}
\sup_{\nu_1 \in \clP_1} \Expect_\lambda^\nu(\tsrp^* -\lambda | \tsrp^* \geq \lambda) = \Expect_\lambda^*(\tsrp^* -\lambda | \tsrp^* \geq \lambda). \label{eqn:srplfddominates}
\end{eqnarray}
We also observe that for any stopping rule $\tau$ that
satisfies the false alarm constraint $\text{FAR}(\tau) \leq
\alpha$, we have,
\begin{eqnarray*}
\begin{aligned}
\sup_{\nu_1 \in \clP_1} \sup_{\lambda \geq 1}\Expect_\lambda^\nu(\tau -\lambda &| \tau \geq \lambda)  \\
&\geq \sup_{\lambda \geq 1}\Expect_\lambda^*(\tau -\lambda | \tau \geq \lambda)  \\
&\geq \sup_{\lambda \geq 1}\Expect_\lambda^*(\tsrp^* -\lambda | \tsrp^* \geq \lambda)  + o(1)\\
&= \sup_{\nu_1 \in \clP_1} \sup_{\lambda \geq 1} \Expect_\lambda^\nu(\tsrp^* -\lambda | \tsrp^* \geq \lambda) + o(1)
\end{aligned}
\end{eqnarray*}
where the second relation follows from the fact that $\tsrp^*$
satisfies the asymptotic optimality of
(\ref{eqn:srpasymproperty}) when the true post-change
distribution is $\underline \nu_1$, and the last equality
follows from (\ref{eqn:srplfddominates}). This completes the
proof of the theorem.
\end{proof}
We note that if the robust SRP stopping rule $\tsrp^*$ is used
when $\clP_0$ is not a singleton, the crucial step of
(\ref{eqn:srpstochorder}) does not hold for $\nu_0 \neq
\overline \nu_0$ and $\nu_1 = \underline \nu_1$. Thus our proof
of optimality of the robust SRP stopping rule does not hold
when the pre-change distribution is unknown.

\subsection{Proof of Theorem \ref{thm:bayesrobust}}

\begin{proof}
The proof is very similar to that of \emph{Case 1} in Theorem
\ref{thm:lordenrobust}. Since the Shiryaev test is optimal for
known distributions, it is clear that the test given in
(\ref{eqn:shiryaevrobust}) is optimal under the Bayesian
criterion when the post-change distribution is $\underline
\nu_1$. Also from the definition of
$\text{PFA}(\tau_{\scriptscriptstyle \mathrm{S}}^*)$ it is
clear that the probability of false alarm depends only on the
pre-change distribution and hence the constraint in (\ref{eqn:
bayescriterionrobust}) is met by the stopping time
$\tau_{\scriptscriptstyle \mathrm{S}}^*$. Hence, it suffices to
show that the value of $\text{ADD}(\tau_{\scriptscriptstyle
\mathrm{S}}^*)$ obtained under any $\nu_1 \in \clP_1$, is no
higher than the value when the true post-change distribution is
$\underline \nu_1$.

Let us first fix $\Lambda = \lambda$. We know by the stochastic
ordering condition that conditioned on $\Lambda = \lambda$, for
all $i \geq \lambda$, we have $\ Y_i^\nu \succ Y_i^*$ where
$Y_i^*$ and $Y_i^\nu$ are as defined in the proof of Theorem
\ref{thm:lordenrobust}. As before, the function,
\[
f'(x_1, \ldots, x_N) \triangleq \max_{1\leq n\leq N }\log \left( \sum_{k = 1}^n \pi_k \exp( \sum_{i=k}^n x_i )  \right)
\]
is continuous and non-decreasing in each of its components as
required by Lemma \ref{lem:incfuncstochord}. Using these facts,
we can show the following by proceeding exactly as in the proof
of Theorem \ref{thm:lordenrobust}: Conditioned on $\Lambda =
\lambda$,
\[
\Expect^*_\lambda ((\tau_{\scriptscriptstyle \mathrm{S}}^* - \lambda)^+ | \clF_{\lambda-1})
\succ \Expect^\nu_\lambda ((\tau_{\scriptscriptstyle \mathrm{S}}^* - \lambda)^+| \clF_{\lambda-1}).
\]
Thus, we have $\Expect^*_\lambda ((\tau_{\scriptscriptstyle
\mathrm{S}}^* - \lambda)^+) \geq \Expect^\nu_\lambda
((\tau_{\scriptscriptstyle \mathrm{S}}^* - \lambda)^+)$ and by
averaging over $\lambda$, we get,
\[
\Expect^*((\tau_{\scriptscriptstyle \mathrm{S}}^* - \Lambda)^+)
\geq \Expect^\nu ((\tau_{\scriptscriptstyle \mathrm{S}}^* - \Lambda)^+).
\]
\end{proof}
\bibliographystyle{IEEEtran}
\bibliography{QCDrefs}
\end{document}